# Active Noise Control Method Using Time Domain Neural Networks for Path Decoupling

Yijing Chu, Qinxuan Xiang, Sipei Zhao, Ming Wu, Y. Zhao, Guangzheng Yu

*Abstract*—In decentralized active noise control (ANC) systems, crosstalk between multichannel secondary sources and error microphones significantly degrades control accuracy. Moreover, prefiltering reference signals in filtered-x (Fx) type algorithms may further introduce modeling errors. A theoretical analysis of the Fx-based decentralized control algorithm was performed, which reveals how prefiltering and crosstalk affect the control performance. Then, a hybrid method combining fixed-value neural networks and adaptive strategies was proposed for efficient decentralized ANC. The adaptive filter models the primary path of its own channel online using the least mean square (LMS) algorithm while the neural network (named DecNet) is used for secondary paths inverting and decoupling. The hybrid DecNet-LMS algorithm was implemented in the time domain to guarantee causality and avoid latency. Simulation results with measured acoustic paths show that the proposed method outperforms the existing ANC algorithms using either traditional adaptive filters or neural network-based fixed-coefficient methods under different acoustic conditions.

## I. Introduction

ACTIVE noise control (ANC) [1]-[3] have been widely used for low-frequency noise cancellation, where the filtered-x (Fx) algorithms [4] have been typically employed due to its stability and efficiency [5][6].

For multichannel ANC (MANC) systems [7][8], the centralized control methods [3] are known for better noise reduction and stability performance at a cost of higher computational complexity [9][10]. As an alternative, the decentralized Fx least mean squares (FxLMS) algorithm [11] has been explored for its computational efficiency [12]. However, the inherent secondary paths coupling could compromise the control accuracy and system stability [13] when one of the channels diverge. Various methods have been explored to tackle this problem, such as the cross-channel compensation approaches [14], various robust algorithms [15][16], and the distributed control methods that balance between the global and local optimizations [17]-[20]. All these strategies cannot prevent crosstalk between different channels, even when a global optimization that employs all the secondary paths is used.

The crosstalk cancellation (CTC) technique with inverse filters [21] was also proposed for MANC, but the tradeoff between causality required by ANC and crosstalk reduction limits their practical applications [22][23].

Recently, neural networks (NNs) have been employed in ANC systems. A convolutional recurrent network (CRN) was proposed to estimate spectra of the cancellation signal directly from the reference signal [24]. Later, a time-domain attentive recurrent network (ARN) with smaller frame sizes was proposed to reduce latency [25]. Other works studied the nonlinear modeling of the primary source [26][27] or increased convergence [28] for correlated noise input. To track moving sources, a hybrid system of an adaptive filter and time-domain CRNs has been proposed using the secondary path-decoupled method, where the CRNs separates the nonlinearity in the secondary source from the linear adaptive filter [29]. Another hybrid system utilized two gated CRNs to model the inverse impulse response of the secondary path such that the adaptive controller simply needs to track the primary path [30]. For MANC systems, the data driven CRN could be extended into multichannel implementation [31][32] without considering the crosstalk between acoustic paths. The crosstalk and secondary path reverse problem was then briefly discussed in [33], where we resorted to nonlinear modeling via NNs in order to reduce the large modeling error.

In this paper, a theoretical analysis was performed to show how the prefiltering in Fx-type algorithms and the crosstalk between multiple secondary paths affect the performance of Fx-based decentralized ANC systems. The analysis reveals that the closed form Wiener solution brings in modeling error in reverberant environments that cannot be well solved by linear models and crosstalk further degrades control performance. To address these issues, a decoupling NN is proposed that solves the reverse and crosstalk in secondary paths. Compared to the conventional inverse FxLMS algorithms that may violate the causality required by ANC or remain large crosstalk, nonlinear modeling could ensure a controllable delay and better CTC. The proposed decoupling network (DecNet) is combined with the LMS algorithm that estimates and tracks the primary path. This method is referred to as the DecNet-LMS. The DecNet was implemented in time domain to mitigate latencies of frequency domain block processing and ensure system causality. Simulations verified the theoretical analysis and showed better convergence and tracking capabilities of the DecNet-LMS than traditional adaptive and existing NN methods.

## II. Performance Analysis of Decentralized Control

### A. Structure of The Decentralized ANC System

A case (1, *K*, *K*) ANC system is shown in Fig. 1, where $K$ loudspeakers, driven by a series of controllers $w_k(n)$, aim to cancel out the primary noise $\{x(n)\}$ at $K$ error microphones. According to the notations in Table I, the primary noise at the

$k$th error microphone can be written as

$$d_k(n) = \boldsymbol{x}_P^T(n)\boldsymbol{p}_k \qquad (1)$$

where $\boldsymbol{x}_P(n) = [x(n), x(n-1), \cdots, x(n-L_P+1)]^T$. Then, the $k$th residual signal can be written as a summation of primary noise, control signals, and background noise, i.e.,

$$e_k(n) = d_k(n) + y_k(n) + \eta_k(n). \qquad (2)$$

The control signals are generated by the loudspeaker driving signals $u_l(n)$ passing through the secondary paths $s_{kl}$, i.e.,

$$y_k(n) = \sum_{l=1}^{K} s_{kl} * u_l(n) = \sum_{l=1}^{K} \boldsymbol{u}_l^T(n)\boldsymbol{s}_{kl} \qquad (3)$$

where "*" is the discrete convolution operator, $\boldsymbol{s}_{kl} = [s_{kl,1} \cdots s_{kl,L_s}]^T$ and $\boldsymbol{u}_l(n) = [u_l(n),...,u_l(n-L_s+1)]^T$.

Assuming the controller changes slowly and the secondary paths are stationary, the convolutions in each channel are commutative. Thus, the control signal can be taken as the reference signal filtered by the secondary path and then the controller [1], which can be written as [34]

$$\begin{aligned}y_k(n) &= \sum_{l=1}^{K}[s_{kl}*x(n)]*w_l(n) = \sum_{l=1}^{K}\boldsymbol{s}_{kl}^T\boldsymbol{x}(n)*w_l(n)\\ &= \sum_{l=1}^{K}\boldsymbol{s}_{kl}^T\boldsymbol{X}(n)\boldsymbol{w}_l(n) = \sum_{l=1}^{K}\boldsymbol{x}_{kl}^T(n)\boldsymbol{w}_l(n),\end{aligned} \qquad (4)$$

where $\boldsymbol{X}(n) = [\boldsymbol{x}(n)\ \boldsymbol{x}(n-1)\ \cdots\ \boldsymbol{x}(n-L+1)]$ is a matrix of size $L_s \times L$, and each column contains the input vector $\boldsymbol{x}(n) = [x(n),...,x(n-L_s+1)]^T$, $\boldsymbol{w}_k(n) = [w_{k,1}(n)...w_{k,L}(n)]^T$ is the $L$-tap controller vector, and $\boldsymbol{x}_{kl}(n) = \boldsymbol{X}^T(n)\boldsymbol{s}_{kl}$ is the filtered input of length $L$. In practice, the secondary paths are unknown and estimated as the secondary path model $\hat{\boldsymbol{s}}_{kl}$. The effect of secondary path modeling error has been analyzed in [35]. To focus on the decoupling problems, we assume that the secondary paths have been well estimated.

### B. Theoretical Analysis of Crosstalk and Prefiltering

In the conventional decentralized control method, the $k$th controller is optimized by minimizing the power of $\{e_k(n)\}$

$$\boldsymbol{w}_{k0} = \arg\min_{\boldsymbol{w}_k} E[e_k^2(n)] = \arg\min_{\boldsymbol{w}_k}(\varepsilon_k + \sum_{l=1,l\neq k}^{K}\tau_{kl}) \qquad (5)$$

where $E[]$ is the expectation operator, $\varepsilon_k$ is related to the $k$th controller, while $\tau_{kl}$ is caused by crosstalk as shown below:

$$\varepsilon_k = E[d_k^2(n)] + 2\boldsymbol{r}_k^T\boldsymbol{w}_k + \boldsymbol{w}_k^T\boldsymbol{R}_{kk}\boldsymbol{w}_k, \qquad (6)$$

$$\tau_{kl} = 2\boldsymbol{w}_k^T\boldsymbol{R}_{kl}\boldsymbol{w}_l, \qquad (7)$$

where $\boldsymbol{r}_k = E[\boldsymbol{x}_{kk}(n)d_k(n)]$ is the correlation vector of length $L$ and $\boldsymbol{R}_{kl} = E[\boldsymbol{x}_{kk}(n)\boldsymbol{x}_{kl}^T(n)]$ is the correlation matrix of size $L$.

In decentralized control methods, $\tau_{kl}$ is assumed to be small and ignored to achieve computational efficiency. Then, the Wiener solution can be obtained from (5) as

$$\boldsymbol{w}_{k0} = -\boldsymbol{R}_{kk}^{-1}\boldsymbol{r}_k. \qquad (8)$$

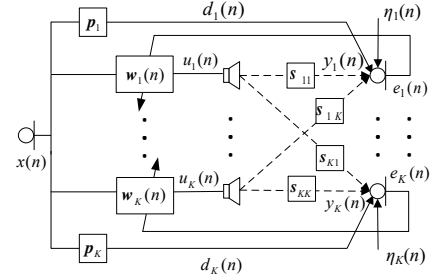

Fig. 1: Block diagram of the $k$th channel of an ANC system using the Fx-based algorithm.

Table I Table of Notations

| | |
|---|---|
| $x(n)$ | primary noise signal at the reference microphone at time $n$ |
| $\boldsymbol{p}_k$ | $L_P$-tap FIR filter from primary noise to the $k$th error microphone |
| $d_k(n)$ | primary noise signal at the $k$th error microphone via $\boldsymbol{p}_k$ at time $n$ |
| $\eta_k(n)$ | Gaussian random noise at the $k$th error microphone at time $n$ |
| $u_k(n)$ | driving signal for the $k$th loudspeaker at time $n$ |
| $\boldsymbol{w}_k(n)$ | $L$-tap FIR control filter for the $k$th secondary source |
| $y_{lk}(n)$ | control sound at the $k$th error microphone from the $l$th loudspeaker |
| $\boldsymbol{s}_{kl}$ | $L_S$-tap FIR filter from the $l$th loudspeaker to the $k$th error microphone |

Substituting this solution to $E[e_k^2(n)]$, the mean square error can be obtained as

$$E[e_k^2(n)] = \varepsilon_k + 2\sum_{l=1,l\neq k}^{K}\boldsymbol{r}_l^T\boldsymbol{R}_{ll}^{-1}\boldsymbol{R}_{kl}\boldsymbol{R}_{kk}^{-1}\boldsymbol{r}_k + \sigma_{\eta,k}^2, \qquad (9)$$

where $\sigma_{\eta,k}^2$ is the variance of the background noise $\{\eta_k(n)\}$, and $\varepsilon_k$ can be further simplified as

$$\begin{aligned}\varepsilon_k &= E[d_k^2(n)] - E[d_k(n)\boldsymbol{x}_{kk}^T(n)]\boldsymbol{R}_{kk}^{-1}E[\boldsymbol{x}_{kk}(n)d_k(n)]\\ &= \boldsymbol{p}_k^T\boldsymbol{R}_P\boldsymbol{p}_k - \boldsymbol{p}_k^T\boldsymbol{R}_{Pk}\boldsymbol{R}_{kk}^{-1}\boldsymbol{R}_{Pk}^T\boldsymbol{p}_k \\ &= Tr[\boldsymbol{R}_P\boldsymbol{p}_k\boldsymbol{p}_k^T] - Tr[\boldsymbol{R}_{Pk}\boldsymbol{R}_{kk}^{-1}\boldsymbol{R}_{Pk}^T\boldsymbol{p}_k\boldsymbol{p}_k^T],\end{aligned} \qquad (10)$$

where $\boldsymbol{R}_P = E[\boldsymbol{x}_P(n)\boldsymbol{x}_P^T(n)]$ and $\boldsymbol{R}_{Pk} = E[\boldsymbol{x}_P(n)\boldsymbol{x}_{kk}^T(n)]$.

**Special Case**: It can be seen that $\boldsymbol{R}_P$ is not necessarily equal to $\boldsymbol{R}_{Pk}\boldsymbol{R}_{kk}^{-1}\boldsymbol{R}_{Pk}^T$, except $s_{kk} = 1$ and $L = L_p$. In this case, we have $\boldsymbol{R}_P = \boldsymbol{R}_{Pk} = \boldsymbol{R}_{kk}$ and $\varepsilon_k = 0$.

To analyze $\varepsilon_k$ under a general condition, $\boldsymbol{x}_P(n)$ and $\boldsymbol{p}_k$ are separated into two parts as $\boldsymbol{x}_P(n) = [\boldsymbol{x}_{0P}^T(n)\ \boldsymbol{x}_{1P}^T(n)]^T$ and $\boldsymbol{p}_k = [\boldsymbol{p}_{0k}^T\ \boldsymbol{p}_{1k}^T]^T$, where $\boldsymbol{x}_{0P}(n)$ and $\boldsymbol{p}_{0k}$ are, respectively, column vectors of the first $L_p - L$ elements of $\boldsymbol{x}_P(n)$ and $\boldsymbol{p}_k$ ($\boldsymbol{x}_P(n)$ and $\boldsymbol{p}_k$ should be patched with zeros if $L > L_p$). Then,

$$\boldsymbol{R}_{Pk} = E[\boldsymbol{x}_P(n)\boldsymbol{x}_{kk}^T(n)] = [\boldsymbol{A}^T\ \boldsymbol{B}^T]^T,$$

where $\boldsymbol{A} = E[\boldsymbol{x}_{0P}(n)\boldsymbol{x}_{kk}^T(n)]$ and $\boldsymbol{B} = E[\boldsymbol{x}_{1P}(n)\boldsymbol{x}_{kk}^T(n)]$.

Consequently, $\boldsymbol{R}_{Pk}\boldsymbol{R}_{kk}^{-1}\boldsymbol{R}_{Pk}^T = \begin{bmatrix} \boldsymbol{A}\boldsymbol{R}_{kk}^{-1}\boldsymbol{A}^T & \boldsymbol{A}\boldsymbol{R}_{kk}^{-1}\boldsymbol{B}^T \\ \boldsymbol{B}\boldsymbol{R}_{kk}^{-1}\boldsymbol{A}^T & \boldsymbol{B}\boldsymbol{R}_{kk}^{-1}\boldsymbol{B}^T \end{bmatrix}$.

In acoustic systems, the primary paths usually have delays for the direct sound to arrive. If $L$ is long enough, $\boldsymbol{p}_{0k} = \boldsymbol{0}$. Then,

$$\varepsilon_k = Tr[\boldsymbol{R}_{kk}\boldsymbol{p}_{1k}\boldsymbol{p}_{1k}^T] - Tr[\boldsymbol{B}\boldsymbol{R}_{kk}^{-1}\boldsymbol{B}^T\boldsymbol{p}_{1k}\boldsymbol{p}_{1k}^T]. \qquad (11)$$

It can be seen that if $s_{kk}$ is an impulse with a delay such that $\boldsymbol{B} = E[\boldsymbol{x}_{1P}(n)\boldsymbol{x}_{kk}^T(n)] = E[\boldsymbol{x}_{kk}(n)\boldsymbol{x}_{kk}^T(n)] = \boldsymbol{R}_{kk}$, then $\varepsilon_k = 0$.

However, Eq. (11) is generally not 0 when the secondary paths are reverberant [36]. The modeling error is even larger [see Eq. (10)] when $L$ is NOT long enough ($p_{0k} \neq 0$). As the filter becomes longer, it is expected that the modeling error no longer decreases but remains at a large value due to the modeling error as shown in (11).

## III. THE PROPOSED DECNET-LMS ALGORITHM

From the above analysis, it can be seen that if $s_{kk}$ has more than one non-zero elements (reverberant environments), the prefiltering in Fx-type algorithms could make $\varepsilon_k \neq 0$. The effect of secondary paths in ANC systems should be modeled as an inverse problem, which cannot generally be solved by linear optimization. On the other hand, crosstalk also brings in extra noise from other channels.

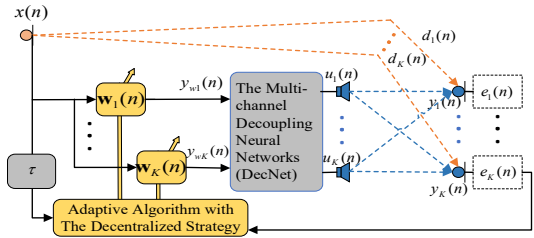

Fig. 2 Diagram of the proposed algorithm for the multichannel ANC system.

To solve these problems, we propose to model the primary and secondary paths separately in the MANC system, as shown in Fig. 2. The DecNet aims to invert the acoustic paths as well as decouple crosstalk in the secondary paths such that the adaptive filters only need to estimate and track the primary paths. Accordingly, the driving signal $u_k(n)$ can be obtained by the following two steps:

1) the reference signal first goes through the adaptive filters $w_k(n)$ to generate the input to the $k$th NN

$$y_{wk}(n) = w_k^T(n) x_L(n), \quad (12)$$

where $x_L(n) = [x(n),...,x(n-L+1)]^T$ is the input of length $L$.

2) then NNs generate driving signals of secondary sources

$$u(n) = \text{DecNet}[Y_w(n)], \quad (13)$$

where $u(n) = [u_1(n) \cdots u_K(n)]^T$, $Y_w(n) = [y_{w1}^T(n) \cdots y_{wK}^T(n)]^T$, $y_{wk}(n) = [y_{wk}(n) \cdots y_{wk}(n-D+1)]^T$, and $D$ is the number of neurons used in the NNs.

Finally, the secondary sound passes through the secondary paths and is collected by the error microphones as

$$y(n) = \begin{bmatrix} y_1(n) \\ \vdots \\ y_K(n) \end{bmatrix} = \begin{bmatrix} s_{11}^T & \cdots & s_{1K}^T \\ \vdots & \ddots & \vdots \\ s_{K1}^T & \cdots & s_{KK}^T \end{bmatrix} \text{vec}[u(n),\cdots,u(n-L_s+1)] \quad (14)$$

where the operator $vec$ columnizes each vector in a matrix.

### A. The Multichannel Decoupling Neural Networks

This section explains how the DecNet is trained to decouple the secondary paths in multichannel ANC systems such that the adaptive controller does not need to prefiltering reference signals and therefore reduces the estimation error.

As shown in Fig. 3, the DecNet (the gray block in Fig. 2) has $K$ NNs, each consisting of $K$ 1-dimension fully connected sub network, namely $nn_{ij}$ ($i,j = 1, ..., K$). The sub network $nn_{ij}$ consists of an input layer of $D$ neurons, a hidden layer with the sigmoid activation function, and an output layer with 1 neuron.

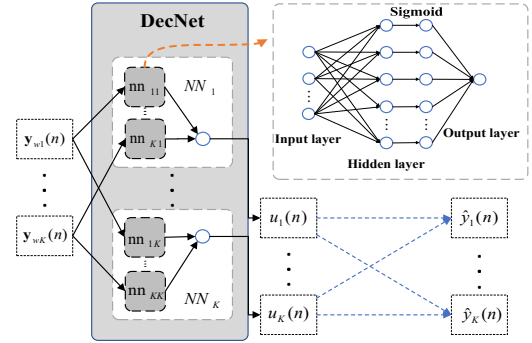

Fig. 3 The structure and training procedure of the Multichannel DecNet.

In the training procedure, the target signal is set as a delayed version of the input signal, i.e. $y_k(n) = y_{wk}(n-\tau)$. The delay should be less than the time delay of the direct sound from the primary source, which ensures causality and enhances prediction accuracy. Then, the mean squared error (MSE) for network training reads

$$L_{NN} = \frac{1}{K} \sum_{l=1}^{K} [y_{wl}(n) - y_l(n)]^2. \quad (15)$$

### B. The Multichannel Adaptive Controller

When the DecNet is well trained, the error microphone receives approximately the delayed input signal

$$y_k(n) \cong y_{wk}(n-\tau) = w_k^T x(n-\tau). \quad (16)$$

Substituting (16) into (2), we obtain

$$e_k(n) = d_k(n) + w_k^T(n) x(n-\tau) + \eta_k(n). \quad (17)$$

It can be seen that the secondary paths have been decoupled by the DecNet such that the adaptive controller does not need to prefilter the secondary path and therefore reduces $\varepsilon_k$ and cancels out $\tau_{kl}$ in the estimation variance. Then, the controller at each channel can be updated by the LMS rule

$$w_k(n+1) = w_k(n) - \mu e_k(n) x(n-\tau). \quad (18)$$

In this case, the Wiener solution to (18) becomes

$$w_{k0} = -R_{xx}^{-1} r_{\Delta k} \quad (19)$$

where $R_{xx} = E[x_L(n) x_L^T(n)]$ is the covariance matrix of the non-filtered input signal $x_L(n)$ and $r_{\Delta k} = E[x_L(n-\tau) d_k(n)]$.

## IV. SIMULATION RESULTS

The modeling error of prefiltering in FxLMS algorithm is evaluated by (10) in this section. The proposed DecNet-LMS algorithm is compared with the decentralized FxLMS (DCFxLMS), the centralized FxLMS (CFxLMS), the Inverse FxLMS [21], and the deep MANC [31] methods.

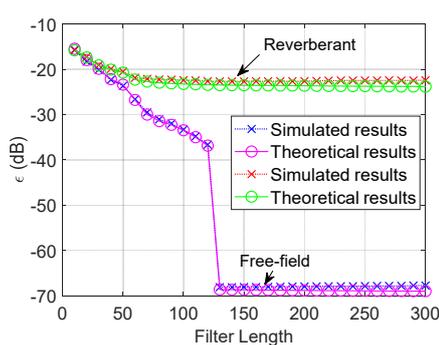 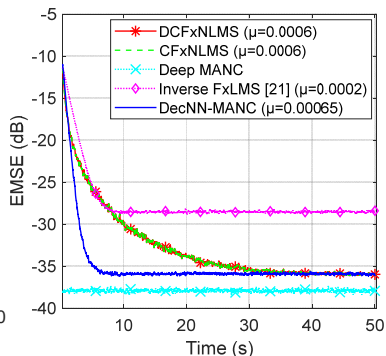 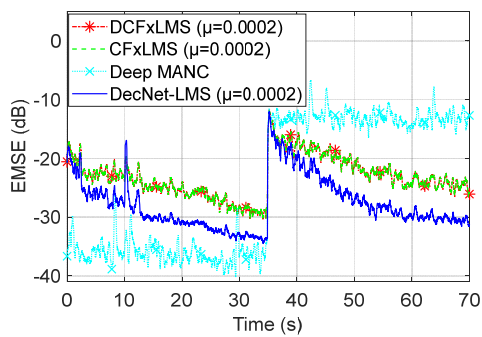

Fig. 4. Evaluation of the theoretical analysis in (10) and (11).

Fig. 5. Averaged EMSE learning curves of different MANC algorithms for white noise. $\mu$ is the step-size for adaptive algorithms.

Fig. 6. EMSE curves of different MANC algorithms for car noise with a time varying primary path. $\mu$ is the step-size for adaptive algorithms.

### A. Experimental Setup

This paper considers a 2-channel ANC system as shown in Fig. 1. The acoustic paths were measured in a room of size 2.80 × 2.30 × 2.10 $m^3$, with an average reverberation time of 0.35 s. The lengths of the primary and secondary paths are 128 and 32 respectively at a sampling rate of 2000 Hz. Each input was segmented into 32-sample frames using a sliding window approach (step size = 1). Each sub network has 32 neurons in the input layer, 512 neurons in the hidden layer, and 1 neuron in the output layer. Under this setting, the complexity of DecNet is 76000, which is higher than the centralized FxLMS with a complexity of 516, but it is much more computationally efficient than the Deep MANC due to the compact model used. It is noted that the DecNet is trained offline in time domain with no additional frequency domain block processing latency.

### B. Effect of Prefiltering

In this experiment, a single channel ANC has been used to evaluate the effect of prefiltering via the modeling error ($\varepsilon$) of the Wiener solution at different lengths. The background noise was set to 0 to eliminate its influence on the modeling errors. The measured reverberant and ideal free-field secondary paths have been tested. The simulated and theoretical modeling errors are plotted in Fig. 4, which shows that the theoretical $\varepsilon$ [Eq. (10), $K = 1$] agrees well with the simulated results. For the free-field case, the modeling error is almost 0 (-70 dB), which agrees with the discussion after (11). However, when the measured reverberant secondary paths were used, both simulation and theoretical results show that the modeling error remains large no matter how long the controller is.

### C. Performance Comparison

Performance of different algorithms are compared in this section with both white noise and car noise as reference signals. For all the methods under test, the length of the adaptive controller was set to $L = 160$.

The averaged excess mean square error (EMSE) curves for white noise are compared in Fig. 5. The proposed DecNet-LMS converges to the steady-state EMSE at the fastest rate compared to the CFxLMS and DCFxLMS, although the fixed controller trained by Deep MANC achieves a lower modeling error by 3 dB. The inverse method in [21] shows increased convergence than CFxLMS since the noise has been whitened, but converges to a large noise power due to difficulties in inverse filter design. When the input signal changed to the car noise, the EMSE curves with only 1 Monte Carlo simulation were presented in Fig. 6. They showed similar convergence performance with that for white noise input during the first 35 seconds. Then, it is assumed that the position of the primary noise changed after 35 seconds, and the tracking capability of these algorithms were further compared. It can be seen that the CRN-based Deep MANC algorithm is not able to track changes in primary paths, while the conventional CFxLMS and DCFxLMS algorithms converge and track the change slowly. By contrast, the proposed DecNet-LMS remains a satisfactory noise reduction performance in this nonstationary case. It is noted that the proposed algorithm has a smaller EMSE than the CFxLMS and DCFxLMS since the car noise is not prefiltered by the secondary paths.

In practical applications, the secondary paths may vary with time, leading to mismatches between the trained DecNet and the true secondary paths. To mitigate potential effects of such mismatches, two methods can be adopted. The first is to employ an adaptive neural network structure for the DecNet to make it track the variations in the secondary paths in real time. However, the real-time training of the adaptive DecNet would lead to higher complexity and cost much more computational resources. The other approach is to account for potential variations in the secondary paths during the training process of DecNet. For example, if the variation is caused by head movement, the secondary paths for several potential head positions can be measured in advance and used for the training of DecNet. If necessary, the coordinates of the head can even be an input to the DecNet and a camera can be used to track the head position in the adaptive control stage. The details of these methods will be investigated in our future work.

### V. CONCLUSION

This paper proposed a DecNet-LMS algorithm for multi-channel active noise control, which addresses the coupling and nonlinearity in secondary path inverse modeling. Simulation results demonstrated that the proposed DecNet-LMS shows better convergence and robustness under various conditions. Future work will investigate the effects of secondary path variation in practical real-time control.